\documentstyle[aaspp4, psfig]{article}

\newcommand{\be}{\begin{equation}}
\newcommand{\ee}{\end{equation}}

\newcommand{\ba}{\begin{eqnarray}}
\newcommand{\ea}{\end{eqnarray}}

\newcommand{\ob}{\omega_B}
\newcommand{\op}{\omega_p}
\newcommand{\om}{\omega}

\begin{document}

\title{Excitation of Alfv\'{e}n Waves and Pulsar Radio Emission}
\author{ Maxim Lyutikov}
\affil{Canadian Institute for Theoretical Astrophysics,
 St George Street,
Toronto, Ontario, M5S 3H8, Canada}
\date{\today}

\begin{abstract}
We analyze  mechanisms of the 
 excitation of Alfv\'{e}n wave  in pulsar magnetospheres as a possible
source of pulsar radio emission generation. 
We find that Cherenkov  excitation of obliquely propagating Alfv\'{e}n waves 
is inefficient, while  excitation at the anomalous 
cyclotron resonance by the particles from the primary beam and 
 from the tail of the bulk distribution function may have a considerable growth
rate. The cyclotron instability on  Alfv\'{e}n waves occurs in the kinetic regime still not
very closed to the star: $r \geq 50 \, R_{NS}$.
We also discuss various  mechanisms of conversion of 
Alfv\'{e}n waves  into escaping radiation. Unfortunately, no  decisive conclusion about
the effectiveness of such conversion can be made.
\end{abstract}

\section{Introduction}

Interpretations of various 
 observational data tend to place the location of radio emission generation
at a distance $r\approx  10-100 R_{NS}$ (e.g., Phillips 1992), though there plenty of claims to the
contrary (Kijak et al. 1999, Smirnova   et al. 1996, Gwinn et al. 1997). As was pointed out by
Kunzl et al. (1998) and Melrose \& Gedalin (1999),
 even with the most conservative  estimates of the 
efficiency of plasma production in the polar caps,  the plasma frequency at those heights  is
much larger that the observed frequency. This argues against radio emission 
mechanisms that generate Langmiur waves with a frequency near the local
plasma frequency (Asseo et al. 1990, Whetheral 1997).
 von Hoensbroech  et al. (1998) argued that
this implies a strong underdense production of particles, but the theoretical 
foundations of such an assumption are weak. Alternatively,
Melrose \& Gedalin (1999)  argued that in order to restrict emission to small
altitudes
the  emission should be  generated at frequencies much smaller than the plasma
frequency, preferably on   Alfv\'{e}n waves. They considered excitation of  
oblique 
 Alfv\'{e}n waves by the Cherenkov resonance with plasma and found that 
it mostly produces waves with $\omega \sim \omega_p$ and thus cannot 
resolve the problem.  Though we agree with their  conclusion that 
Cherenkov  excitation of Alfv\'{e}n waves is insignificant in the pulsar magnetosphere,
we disagree on the reasons why. First, 
we do not agree with the 
conclusion of Melrose \& Gedalin (1999)   that Cherenkov resonance of the beam particles
with the Alfv\'{e}n waves occurs outside the light cylinder for the 
conventional beam energies. It actually occurs for radii $r \leq 50 R_{NS}$ 
(see section \ref{waves}). 
Secondly, they assumed that Cherenkov  excitation of Alfv\'{e}n waves 
occurs in a kinetic regime, while it was shown in Lyutikov (1999) that
Cherenkov  excitation of Alfv\'{e}n waves 
occurs in a hydrodynamic regime.

On the other hand, some authors postulated excitation of the Alfv\'{e}n mode and,
using the fact that Alfv\'{e}n is guided along the curved magnetic field, were
able to explain some observational data like dependence of
 the mean profile on the frequency (e.g. Barnard \& Arons 1986, McKinnon 1994, Gallant 1998,
Gwinn et al. 1999).  
The above arguments stimulated us to reconsider the possibility of 
excitation  of Alfv\'{e}n  waves at the low altitudes in the magnetosphere,
{\it including} the excitation of waves at 
anomalous cyclotron resonance.  

A major theoretical problem of the theories that produce radio emission on Alfv\'{e}n
waves is that 
Alfv\'{e}n waves cannot escape from the plasma and have to be converted into
escaping radiation (preferably X mode). In section \ref{conversion} we review
various possibilities of
Alfv\'{e}n waves conversion into escaping modes.

\section{Waves and resonances}
\label{waves}

 The open field lines  of the pulsar magnetosphere are 
 populated by the dense one-dimensional
flow of electron-positron
pair plasma penetrated by highly  energetic primary beam with the
density equal to the Goldreich-Julian density $n_{GJ}={\bf \Omega. B}/( 2\,
\pi\, e\, c)$, Lorentz factor $\gamma_b \approx 10^6$ 
(Arons 1983,  Daugherty  \& Harding  1996). The density of the pair plasma is $n
\approx \lambda _M \, n_{GJ} =10^3-10^5 n_{GJ}$, where $\lambda_M$ is the multiplicity
factor which gives the number of pairs produced by each primary particle; its
Lorentz factor is $\gamma_p \approx \gamma_b/ \lambda_M=10-10^3$.
The distribution function of the bulk plasma also has a high energy tail up to
to the Lorentz factor $\gamma_t \approx 10^5$.

Though the pulsar plasma is thought to be relativistically hot, 
we restrict ourselves to cold plasma case which simplifies consideration
considerably. Except for the Landau damping of the waves with slow phase velocity,
thermal effects has only marginal importance for the wave-particles interaction and
growth rates of instabilities
 (Lyutikov 1999a). In what follows
 we consider wave excitation  in the plasma frame and  then use the transformation 
rules of the Lorentz factors and  frequencies from the plasma frame to 
pulsar frame:
$\gamma' = 2 \gamma_p \gamma$, $\om_p'= \sqrt{ \gamma_p} \om_p$,
$\ob ' =\ob$, where primes denote quantities in the pulsar frame
and $\gamma_p$ is the Lorentz factor of the relative motion of the plasma frame to the
pulsar frame. 

 In the strongly  magnetized  pair plasma,  in the small frequency limit
$\omega \ll \omega_p$, there  are two modes - transverse extraordinary (X) mode, with the
electric vector perpendicular to the {\bf k-B} plane,
and quasitransverse  Alfv\'{e}n wave  with
 the electric vector in the {\bf k-B} plane. 
 In the
plasma frame 
 the dispersion relations of the extraordinary  and Alfv\'{e}n  modes 
are (e.g. Lyutikov et al. 1999a).
\ba
&&
\omega_A =
k c \cos \theta \left( 1 - {{{{{\it  \omega_p}}^2}}\over {{{{\it  {\omega_B}}}^2}}} -
    {{{ k^2 c^2 }\,{{\sin  ^2 \theta}}}\over {4\,{{{\it  \omega_p}}^2}}} \right), 
\hskip .3 truein \mbox{ for $\omega \ll \omega_p$}
\label{ss}
\\ &&
\omega_X= k c \left( 1- { \omega_p^2 \over \omega_B^2} \right)
\label{ss2}
\ea

The   plasma normal modes  can be excited  due to interaction with particles at 
 the  Cherenkov 
\be 
\omega - k_{\parallel} v_{\parallel} =0
\label{ss3}
\ee
 and anomalous cyclotron resonances
\be 
\omega - k_{\parallel} v_{\parallel} + \omega_B/\gamma_{res} =0
\label{ss1}
\ee
where $\omega $ is a frequency of the wave, $k_{\parallel}$ and $ v_{\parallel}$ are  projections
of the wave vector and velocity along the direction of the magnetic field, 
$\omega_B$ is a (nonrelativistic) cyclotron frequency and $\gamma_{res}$ is a Lorentz factor
of the particles. Also 
note plus sign in front of the cyclotron term in Eq. (\ref{ss1}). 
Two regimes of excitaion are possible:
kinetic and hydrodynamic.
\footnote{
In a hydrodynamic type  instability all the
particles of the beam resonate with one normal wave in the plasma,
while in a kinetic regime only small fraction of the beam 
particles  resonates with  a given wave.}

The X mode can not be excited by the Cherenkov-type
interaction since it has has $ {\bf E } \perp  {\bf B}, {\bf v} $.
At radii larger than $\approx 50 R_{NS}$ (see Eq. \ref{uiaj22}) the phase
velocity  of the of the X mode becomes smaller than the velocity of the 
primary beam with $\gamma_b \approx 10^6$,  
so it can in principle be excited by cyclotron
resonance, but near the surface of the neutron star  the growth rate
of the cyclotron excitation of the X mode is extremely small 
and the  frequencies do not correspond to observed ones (Lyutikov 1999a). 
The cyclotron excitation of X mode
may occur in the outer region of pulsar magnetosphere and is considered
as a viable mechanism of pulsar radio emission generation  (Kazbegi et al. 1989, Machabeli \& Usov 1989,
Lyutikov et al. 1999b). Thus,  we conclude that near the surface the X mode cannot be excited 
and  in what follows we concentrate on the possible excitation of the  Alfv\'{e}n mode. 

\section{Cherenkov excitation of the Alfv\'{e}n mode}

The   Cherenkov  excitation of Alfv\'{e}n waves has been 
considered  in detail by Lyutikov (1999a). On a microphysical scale
it is similar to the Cherenkov excitation of the Langmuir waves - 
the motion of the resonant particle is coupled to the component of
the electric field
of the wave along the magnetic field (and the velocity of the particle). 
It  is possible only
for oblique propagation since Alfv\'{e}n waves which have a component of electric field 
parallel to  the external magnetic field $e_z \approx \sin \theta$. 

Using resonance condition (\ref{ss3}) and the 
 low frequency asymptotics of Alfv\'{e}n waves (\ref{ss}) we infer that
that the possibility of the
 Cherenkov excitation of the Alfv\'{e}n waves by the particles from the
primary beam depends on the parameter
\begin{equation}
\mu\,=\,
{ 2 \,\gamma_b \, \omega_p \over {\omega_B}} = 2^{3/2}
 \gamma_b \sqrt{ { \lambda \, \Omega \over    {\omega_B}} }
=\, 2^{3/2}\,
\sqrt{ { \lambda \, \Omega \over  {\omega_B}} } { \gamma_b^{\prime} \over 
\gamma_p ^{3/2} } \,=\, 
5 \times 10^{-3} \left( {r\over R_{NS}} \right)^{3/2}  \,=\,
\left\{ \begin{array}{ll}
< 1, & \mbox{ if $  \left( {r\over R_{NS}} \right) \leq 50 $} \\
> 1,  & \mbox{ if $  \left( {r\over R_{NS}} \right) \geq  50 $}
\end{array}\right.
\label{uiaj22}
\end{equation}

\begin{figure}
\psfig{file=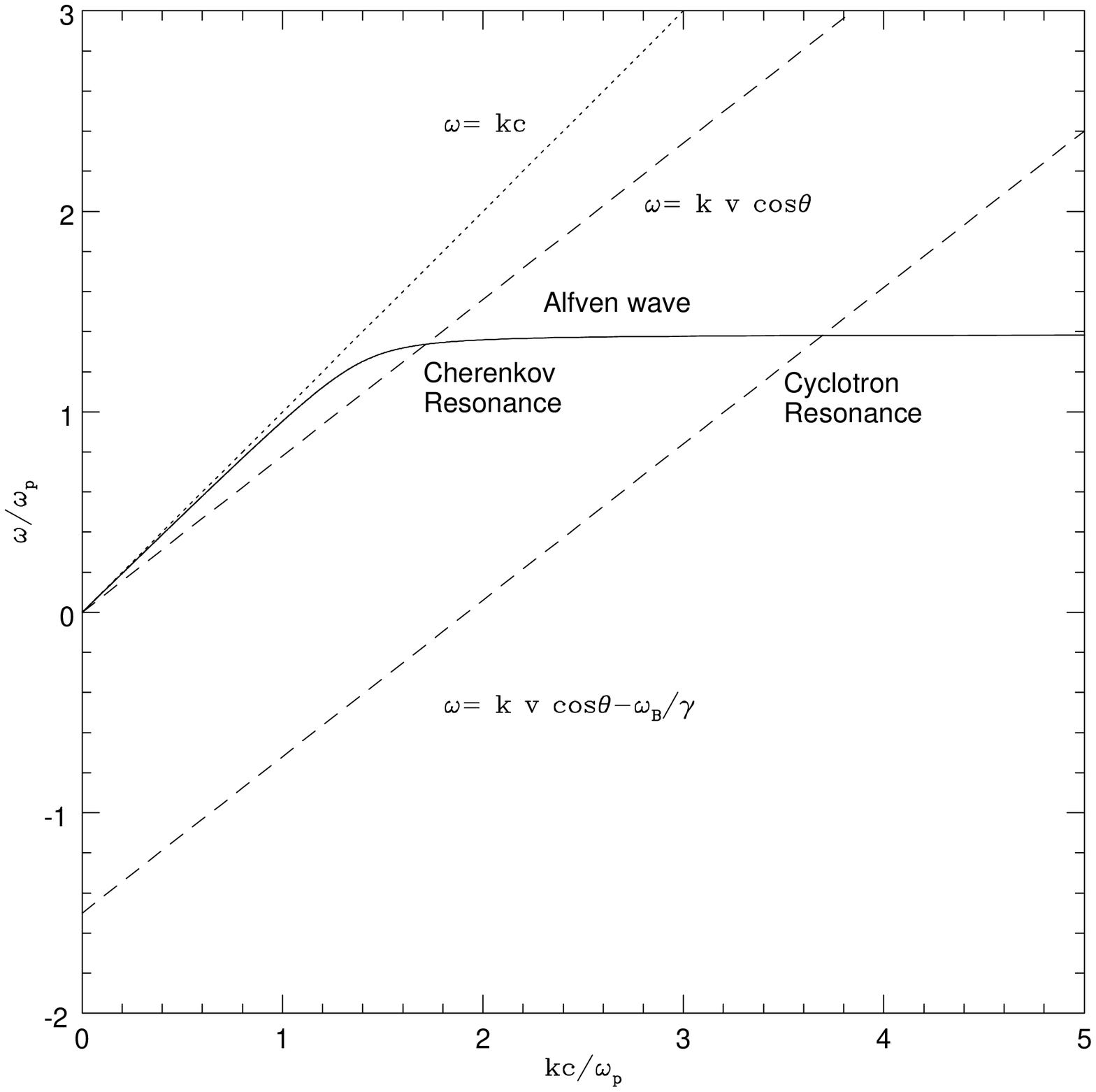,width=15.0cm}
\label{fig2}
\end{figure}

\begin{figure}
\psfig{file=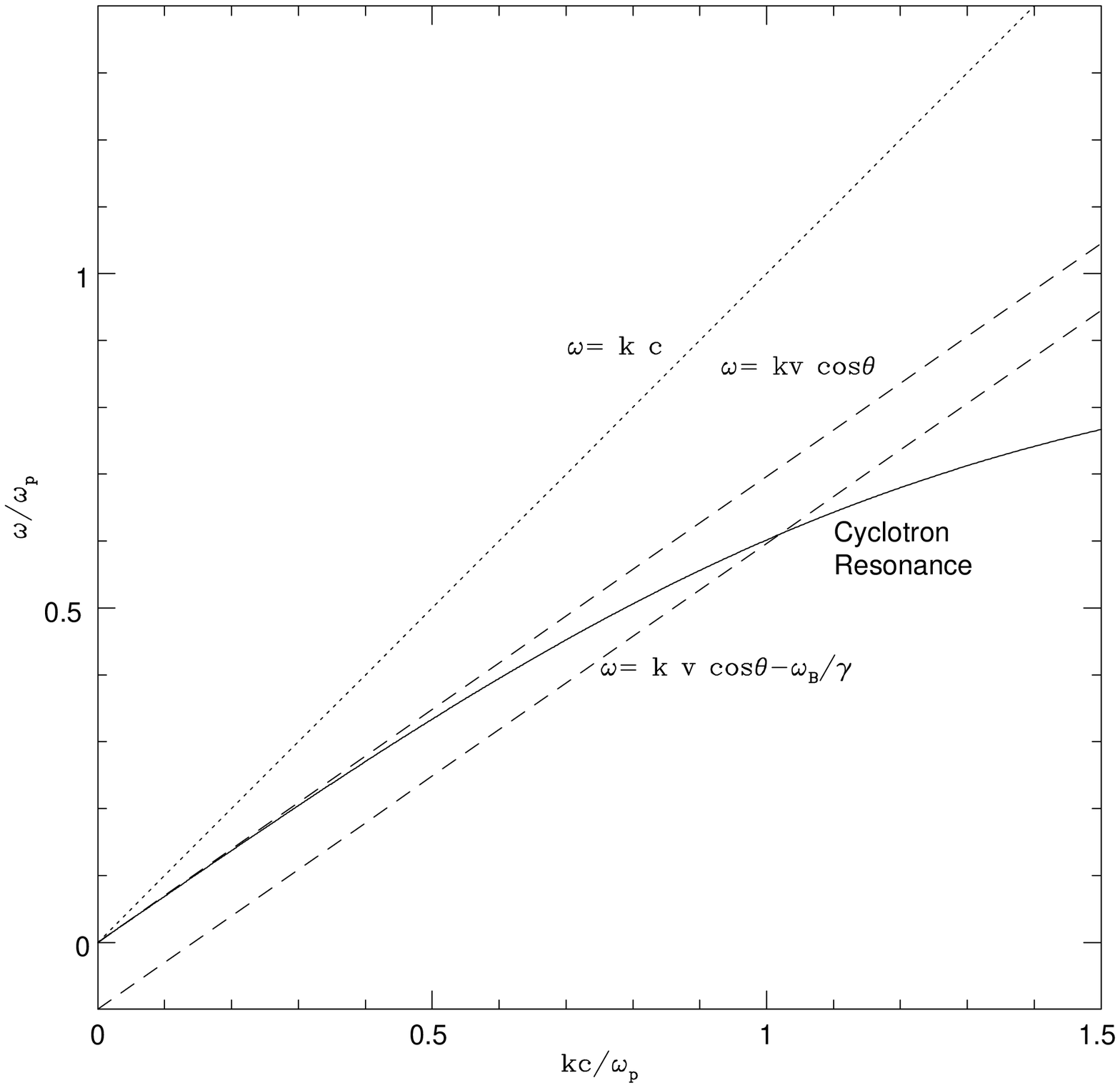,width=15.0cm}
\label{fig3}
\end{figure}

Alfv\'{e}n waves can be excited by Cherenkov
resonance only for  $\mu \, < \, 1$, when the phase velocity of the 
fast particles may become equal to the phase velocity of obliquely propagating 
Alfv\'{e}n waves  (see Figs. \ref{fig2} and \ref{fig3}). 
When $\mu \, >\, 1$ the particles move faster that the beam and,
due to the specific  dispersion of  Alfv\'{e}n waves,   cannot  resonate with them. 
In the outer parts of magnetosphere ( $ r \geq 50 R_{NS} $)
parameter $\mu $ becomes much larger than unity
$\mu \gg 1$  so Alfv\'{e}n waves cannot be excited by  Cherenkov
interaction. 

The resonant condition for the Cherenkov excitation of Alfv\'{e}n waves
(Eq. \ref{ss3}) may be solved 
for $k_{\perp}= k \sin \theta $:
\be
k_{\perp}= \omega_p \sqrt{ {1\over 2  \gamma_b^2} -  {\omega_p^2 \over \omega_B^2} }
\ee
As expected,
 excitation is limited to ${\omega_p^2 \over \omega_B^2}  <{1\over 2  \gamma_b^2}$, e.i. for 
 $r \leq 50 R_{NS}$. 

The growth rate for the Cherenkov excitation of Alfv\'{e}n waves
(which occurs in the hydrodynamic regime)
has been calculated in Lyutikov (1999) (see also Godfrey et al. 1975):
\begin{equation}
\Delta =  
  {\frac{{\sqrt{3}}\,{\omega_p^{{\frac{1}{3}}}}\,
       {{\omega_{GJ}}^{{\frac{2}{3}}}}\,\cot \theta}{
       {2^{{\frac{7}{6}}}}\,{\gamma_b^{{\frac{8}{3}}}}}} 
\label{fhdur2}
\end{equation}
where $\omega_{GJ} ^2 = 4 \pi e^2 n_{GJ} /m$ is the plasma frequency associated with 
the Goldreich-Julian density.
It is very similar to the growth of the Cherenkov excitation of Langmuir  waves. 
This result is expected since the microphysics of the Cherenkov excitation
of Langmuir and Alfv\'{e}n waves is the same - coupling of the parallel (to the 
magnetic field) motion of 
particles to the parallel component of the electric field of the wave. 
Consequently, 
the growth rate  given by Eq. (\ref{fhdur2}) for the Alfv\'{e}n wave excitation
 suffers from the same problem as Langmuir wave excitation: it is strongly suppressed by 
large Lorentz factor of the primary beam. 
We thus conclude that Cherenkov excitation of the Alfv\'{e}n waves is ineffective.

\section{Cyclotron excitation of the Alfv\'{e}n mode}

The other possibility of excitation of Alfv\'{e}n waves is by
the anomalous  cyclotron resonance
(Tsytovich \& Kaplan 1972, Hardee \& Rose 1978, Lyutikov 1999a). 
On a microphysical scale, during an emission at the
 anomalous  cyclotron resonance, 
a particle undergoes a transition {\it up}
in Landau levels coupling  transverse velocity to the electric field
of the wave (Ginzburg  \& Eidman  1959). 

In case of low frequency ($\omega_A \ll \omega_p$) waves, the 
the resonance condition for the cyclotron excitation of Alfv\'{e}n reads
\begin{equation}
k_{res}
c \, \cos\theta\,\left( {1\over 2 \gamma_b^2}-\, {\omega_p^2\over {\omega_B}^2}
-\, {k^2\, c^2 \, \sin^2 \theta  \over 4 \omega_p^2} \right) +
{\omega_B}/\gamma_b=\, 0,
\hskip .2 truein \mbox{ $ k_{res} c \ll \omega_p$}
\label{fj3}
\end{equation}
If  the third term is much larger
than the first two  (this happens for the  angles of propagation larger than some
 critical angles  ${ \omega_p \over  \gamma_b \omega}$
 and $\,{\gamma_b \, \omega_p^4\over {\omega_B}^4}
$ ) the resonance occurs at
\begin{equation}
k_{res} c= \left( { 4 \, \omega_p^2 \, {\omega_B} \over \gamma_b\, \cos \theta\,
\sin ^2 \theta} \right)^{1/3} \approx {  \omega_p  \over \mu^{1/3} }
  \left( {1 \over \cos \theta\,
\sin ^2 \theta} \right)^{1/3} 
\label{fj4}
\end{equation}
The condition $ \omega_{res} \ll \omega_p$, which
guarantees that Alfv\'{e}n waves are not
damped by the Cherenkov interaction with the bulk plasma particles,  requires
 $\mu \gg   {1 \over \cos \theta\,
\sin ^2 \theta  }  $, or equivalently, 
$r \geq 50 R_{NS}$.  
This is a serious restriction: Alfv\'{e}n waves cannot be excited at lower
altitudes since then the cyclotron resonance occurs in the region where
Alfv\'{e}n waves  are strongly damped due to Cherenkov resonance with the
bulk particles (Arons \& Barnard 1986). 

Cyclotron excitation of Alfv\'{e}n waves occurs in  kinetic regime 
(Lyutikov 1999a).
The growth rate is 
\begin{equation}
\Gamma = {\pi \over 4} \, {\omega_b ^2 \over \omega_{res} \Delta \gamma} 
\label{piajd24}
\end{equation}
where $\Delta \gamma$ is the scatter in Lorentz factors of the resonant particles and 
 the resonant frequency  $\omega_{res}$ follows from Eq. (\ref{fj4}). 
Formally,  growth rate for the cyclotron instability on Alfv\'{e}n waves 
(Eq (\ref{piajd24})) is the same
 as the growth rate for the the cyclotron instability on the high frequency 
transverse waves (Lyutikov et al. 1999b, Kazbegi  et al. 1989),  which occurs in the outer
parts of the pulsar magnetosphere. The important difference is that 
the cyclotron instability on Alfv\'{e}n waves 
can occur at lower altitudes, where the density of the resonant particles is higher.
\footnote{In the case of Alfv\'{e}n waves we use the dispersive correction
$k_{\perp}^2  c^2 \sin^2 \theta / \omega_p^2$ instead of $\om_p^2/\ob^2$.}

Numerically, for the emission generated at $r\approx 50 R _{NS}$ 
and a quite narrow primary beam ($\Delta \gamma \approx 10^2$ - see Lyutikov et al. 1999a), we
have
\be 
\Gamma \approx 
10^ 5 {\rm sec^{-1}}
\ee
Thus, we conclude that the growth rate of the 
 cyclotron instability  on the Alfv\'{e}n waves  may be considerable enough to account for the
high brightness pulsar radio emission.

\section{Wave conversion}
\label{conversion}

There are two  fundamental problem with Alfv\'{e}n waves - they cannot escape from plasma
and  they are damped on the Cherenkov resonance with 
plasma particles when their frequency becomes comparable to $\omega_p$. 
As  Alfv\'{e}n waves propagate into decreasing plasma density, their
frequency will eventually become comparable to the plasma frequency
(Barnard \& Arons 1986). Before that, Alfv\'{e}n waves have to be  converted
 into escaping modes. 
The conversion should take place at  such radii that the local plasma
frequency, transformed into the pulsar frame, is still larger than the observed frequencies:
$ r \leq 500 R_{NS}$. 

There are two generic types of conversion - linear and nonlinear. 
Linear conversion occurs in the regions where WKB approximation 
for wave propagation is not satisfied (e.g., Zheleznyakov 1996). Nonlinear 
conversion is due to the wave-wave or wave-particles interaction. 
In order to escape absorption at the Cherenkov resonance the 
Alfv\'{e}n waves should converted into either superluminous ordinary waves
or into X modes that does not have a component of the electric field along
the external magnetic field. 

\subsection{Linear  conversion} 

Linear conversion of waves occurs when  dispersion curves 
of two modes
approach each other closely on the $\om - k$
diagram.
Effective conversion occurs when the distance between two dispersion curves
 becomes comparable to the "width" of the
dispersion curve. Several process can contribute to the broadening of the
dispersion curve. First, the  inhomogeneity of the medium induces a width 
${\delta \omega \over \omega} \approx {1  \over  k L}$ where $L$ is 
a typical inhomogeneity scale. Inhomogeneities can be due both to 
large scale (of the order of the light cylinder radius) 
 density  fluctuations, excited by temporal and 
special modulations of the flow, and due to the small scale plasma turbulence. 
Linear conversion of 
Alfv\'{e}n waves into X mode is impossible because of their different 
polarizations. Linear conversion of 
Alfv\'{e}n waves into O mode can occur only when the frequency of the 
Alfv\'{e}n waves approaches plasma frequency, e.i. exactly at the moment when
Alfv\'{e}n waves  become strongly damped at the Cherenkov resonance with the
bulk plasma. The effectiveness of the conversion then depends on the not
very well known details of the bulk plasma distribution function (e.g., distributions with
considerable high energy tail will tend to damp the waves stronger), so no decisive
argument  about the effectiveness of such conversion can  be made.

Secondly, the wave turbulence results in effective collisions (with frequency 
$\nu_c$)
of Alfv\'{e}n waves with each other; in this case  the  induced width is 
 ${\delta \omega } \approx {\nu_c }$. \footnote{Terminology here may be 
a bit confusing: the frequency of collisions, of course, depends on the total energy
density of turbulence.}
 For a given turbulent energy density  $W_{tot}$ the typical wave-wave collision
frequency (based on a three wave interaction) can be estimated as 
\be 
\nu_c \approx \left( {e \over m c} \right)^2   {W_{tot}  \over  \omega_p  }
\ee
Effective conversion occurs when this  frequency  is of the order of the minimum
difference between the dispersion curves. In the case of strong turbulence the
 the collision frequency becomes comparable to the frequency of the waves near the
conversion point: $\nu \approx \omega_p$.  In that case
the required wave energy density is  approximately equal to the energy in plasma.
Assuming that the energy density of plasma is of the order of the thermal energy, 
we find
\be 
W_{tot} 
= \left( {\om_p mc \over e } \right)^2 \approx n mc^2
\label{12}
\ee
In addition,
in the pulsar plasma the energy in the secondary plasma is approximately equal to the
energy in the beam. Then, the 
energy density given by Eq. (\ref{12}) is of the order  of the beam energy. 
The amount of energy lost by a beam due to the wave excitation depends
on the complicated physics of the instability  saturation mechanisms, but 
we can reasonably expect that  to be $\sim 1-10 \%$. In this case the
 the collisional  frequency $\nu_c \sim 0.01 - 0.1 \op$ and the
 collisional conversion of Alfv\'{e}n  waves into O mode 
can be marginally effective.

Another mechanism of linear wave transformation is related to the presence
of velocity shear in the flow (Arons \& Smith 1979,  Mohajan et al. 1997, 
 Chagelishvili et al. 1997). 
We can reasonably expect that the flow of pair plasma 
has some shear associated with the plasma generation in the 
polar gap. Given a strong shear Langmuir waves become coupled 
 to the  escaping O modes.
 Whether the shear expected in the polar outflow can provide
enough coupling  remains uncertain.

\subsection{Nonlinear conversion}

\subsubsection{Wave - wave interaction}

Consider a merger of two Alfv\'{e}n waves into escaping X or O mode. 
For the three wave processes to take place, the participating waves
 should satisfy resonance conditions (conservation of energy and
momentum)
and more subtle
conditions on the polarization determined by the
matrix elements of the  third order nonlinear current (Melrose
1978 Eqs. (10.105) and (10.125)). From the energy and momentum conservation it is 
easy to see that only oppositely propagating Alfv\'{e}n waves can merge into X or O mode.

In calculating the matrix element of the three-wave interaction
two simplifications are possible in the case of strongly  magnetized electron-positron plasma.
First, nonlinear current terms which are proportional to an odd power of the sign of the charge
will cancel out. Secondly, we can make an expansion in powers  $1/\ob$ 
and keep the lowest order. Under these assumption 
the matrix element becomes
\be 
V \approx - { i e \over m c} {\omega_p \over \omega_B}
\label{WW}
\ee
Given the matrix element (\ref{WW}) we can find the the probability of emission
in the random phase approximation (which assumes that radiation is broadband)
 (Melrose 1978)
\be
u \approx   V^2   \omega _p = {e^2 \over m^2  c^2} {\omega_p ^3 \over \omega_B^2}
\ee
Then, again assuming that the energy density of plasma is of the order of the thermal energy,
the characteristic nonlinear decay  time  is
\be 
\Gamma \approx  {\omega_p ^3 \over \omega_B^2} = 2^{3/2} \Omega 
 \sqrt{ {\lambda_M^3 \Omega \over \omega_B} } \approx 2 \times 10^{-2} 
\left( { r\over R_{NS}} \right)^{3/2}
\ee
(compare with Mikhailovskii 1980).
So that at the distance $r \approx 50-100 R_{NS}$ the nonlinear conversion of
Alfv\'{e}n waves is only marginally possible. The nonlinear conversion of Alfv\'{e}n waves
at lower altitudes is impossible because of the small growth rate.

Thus, we come to a conclusion that though in principle the Alfv\'{e}n waves can be
converted into escaping modes by nonlinear selfinteraction, the conversion 
 must occur at comparatively large
radii $r \approx 50-100 R_{NS}$. There is another 
 serious  problem with nonlinear
conversion - it requires  a presence of strong backward propagating waves. 
Presence of such  backward propagating  waves is not obvious in the 
pulsar magnetosphere. They should either be excited by a backward 
propagating fast particles,  whose origin cannot be simply justified, or be a 
result of a scattering of   the initial 
forwards propagating  waves. The Thompson scattering is strongly 
suppressed in the magnetized plasma and is probably ineffective, but 
induced scattering (see below) may provide such backward propagating waves.

\subsubsection{Induced scattering}

Induced scattering of longitudinal waves in pulsar magnetosphere
have been 
considered by Machabeli (1983)  and 
Lyubarsky \& Petrova (1996). They showed that  it  may be an effective
process of transferring energy from the Alfv\'{e}n branch to the O mode.
There are several problems with this mechanism. First, the effective
induced scattering of Alfv\'{e}n waves on plasma particles 
in a superstrong magnetic field  occurs in the
same region where waves become strongly damped, the question which process
is more effective (scattering or damping) again strongly depends on the unknown details
of the plasma distribution function. Secondly, if scattering is effective,
the  induced  scattering transfers energy to
smallest wave vectors of the O mode (Langmiur condensate).  
Propagation and escape of the  O mode has not been properly investigated, but since
the  O mode with initially small wave vector  has a very small index of refraction
($n \sim 0$) it will be strongly refracted (by the angle $\approx \pi$)
as it converts into a vacuum mode with   ($n \sim 1$). 
This strong refraction  would then contradict the observed narrow pulsar profiles.

Summarizing this section, 
 we conclude that at this point  we  cannot  make a decisive statement
whether   linear or nonlinear conversion  of 
 Alfv\'{e}n wave  into escaping modes is effective.

\section{Conclusion}

The results of this work confirm the conclusion of 
 Lyutikov (1999a) that in the pulsar magnetosphere the
 electromagnetic  cyclotron instabilities  are the most likely candidates for
the pulsar radio emission generation. These instabilities develop on the   X and O modes
 (in the outer regions of the pulsar
magnetosphere) and on the   Alfv\'{e}n mode (possibly in the lower,
$r \sim 50-100 R_{NS}$, regions).
Cyclotron instabilities on the X and O modes 
have  smaller (than on the Alfv\'{e}n waves)
growth rate, but generate waves that can directly escape 
from the plasma. The  growth rates of the  cyclotron
 instability  on Alfv\'{e}n waves can be very large, 
but  the complications of the  wave conversion and 
 absorption in the outflowing plasma, which depend on the 
unconstrained details of the
plasma distribution function, may 
put the model based
on the  Alfv\'{e}n wave excitation at disadvantage.

Both excitation and the nonlinear conversion of Alfv\'{e}n waves is possible only
for $r \geq  50-100 R_{NS}$.  
This is a comparatively large altitudes. If we  suppose that the opening angle 
of the magnetic field lines controls the width of the pulsar emission beam (e.g, Rankin 1992), then
the value of the opening angle  would imply emission right near the surface $r \sim R_{NS}$.
If taken at a face value, this interpretation would exclude the  Alfv\'{e}n waves as
a source of observed radio emission.

The comparatively large radii of emission, $r \geq  50-100 R_{NS}$ and larger, may not be that unacceptable
from the observational point of view as well (Lyutikov 1999b). Such effects as 
"wide beam" geometry (Manchester 1995), emission bridge between some widely separated pulses,
extra peaks in Crab at  high frequencies maybe naturally explained by large 
altitudes of emission. The results of this work  stress one again the difficulties of 
wave excitation at very small altitudes:  Alfv\'{e}n waves, as well as ordinary
and extraordinary modes, cannot be excited  beam instabilities or converted nonlinearly into escaping modes 
at low altitudes. On the other hand, both excitation and conversion should occur
at larger $r \geq 50 R_{NS}$ altitudes which remain our prefered regions 
for the generation of pulsar radio emission.

\acknowledgements
I would like to thank George Machabeli for discussions and  useful comments.

\newpage
FIGURE CAPTIONS

Fig. 1(a). Resonances on the Alfv\'{e}n mode  for $\mu < 1$.

Fig. 1(b). Resonances on the Alfv\'{e}n mode for $\mu> 1$.
\newpage

%file ~lyutikov/pulsar/macroHydroCyclo  ; macro Alfvenres1

\end{document}